\documentclass[12pt,a4paper,english,aps,superscriptaddress,groupedaddress,showpacs,showkeys]{revtex4}

\usepackage[T1]{fontenc}
\usepackage[latin9]{inputenc}
\usepackage{babel}
\usepackage{textcomp}
\usepackage{amsmath}
\usepackage{amssymb}
\usepackage{graphicx}
\usepackage[unicode=true]
 {hyperref}
\usepackage{breakurl}

\makeatletter

\@ifundefined{textcolor}{}
{%
 \definecolor{BLACK}{gray}{0}
 \definecolor{WHITE}{gray}{1}
 \definecolor{RED}{rgb}{1,0,0}
 \definecolor{GREEN}{rgb}{0,1,0}
 \definecolor{BLUE}{rgb}{0,0,1}
 \definecolor{CYAN}{cmyk}{1,0,0,0}
 \definecolor{MAGENTA}{cmyk}{0,1,0,0}
 \definecolor{YELLOW}{cmyk}{0,0,1,0}
}

\usepackage{dcolumn}
\usepackage{bm}
\usepackage{amsfonts}

\makeatother

\begin{document}

\title{On the locus formed by the maximum heights of an ultra-relativistic
projectile}

\author{Salvatore De Vincenzo}

\homepage{https://orcid.org/0000-0002-5009-053X}

\email{[salvatored@nu.ac.th]}

\selectlanguage{english}%

\affiliation{The Institute for Fundamental Study (IF), Naresuan University, Phitsanulok
65000, Thailand}

\date{January 13, 2025}
\begin{abstract}
\noindent We consider the problem of relativistic projectiles in a
uniform gravitational force field in an inertial frame. For the first
time, we have found the curve that joins the points of maximum height
of all trajectories followed by a projectile in the ultra-relativistic
limit. The parametric equations of this curve produce an onion-like
curve; in fact, it is one of the loops of a lemniscate-type curve.
We also verify that the curve is an ellipse in the nonrelativistic
approximation. These two limiting results are obtained by following
two slightly distinct approaches. In addition, we calculate the nonrelativistic
and ultra-relativistic approximations of the trajectory equation and
parametric equations of the trajectory as functions of time. We also
find the asymptotic behavior of the ultra-relativistic trajectory
equations for points that are very distant from the launching point.
All limiting cases in the article are studied in detail. Additionally,
we discuss various physical issues related to some of our results.
The content of the article is appropriate for advanced undergraduate
students. 
\end{abstract}



\maketitle

\section{Introduction}

\noindent Consider a nonrelativistic projectile that is fired from
the ground with a launch angle $\theta$ and initial speed $v_{0}$
in a uniform gravitational force field and in the absence of air (i.e.,
no drag force). If one places the origin $\mathrm{O}$ of the coordinate
system at the launch point and makes $v_{0}=\mathrm{constant}$ but
allows $\theta$ to vary between $\theta=0\text{\textdegree}$ and
$\theta=180\text{\textdegree}$, one obtains different parabolic trajectories
that share an unexpected property: all points of the maximum height
of these parabolas lie on an ellipse whose eccentricity is $\epsilon=\sqrt{3}/2$.
This result was first mentioned in an old mechanics book by MacMillan
\cite{RefA}. Many years later, Fern\'{a}ndez-Chapou \textit{et al}.
rediscovered this property \cite{RefB}. Since then, several papers
have presented and analyzed this topic. \cite{RefC,RefD,RefE,RefF,RefG,RefH,RefI}.
Alternatively, if a drag force opposite to the projectile velocity
is added to the problem, i.e., $\mathbf{f}=-\mathrm{m}b\,\mathbf{v}$
($\mathrm{m}$ is the rest mass of the projectile, and $b$ is the
drag coefficient), the points of maximum height of all trajectories
followed by the projectile lie on a Lambert $W$ function \cite{RefJ,RefK}.
Certainly, the latter curve becomes an ellipse when $b\rightarrow0$,
as expected \cite{RefJ}. 

Different interesting papers address the problem of a relativistic
projectile in a uniform force field (electric or gravitational) in
different manners \cite{RefL,RefM,RefN,RefO,RefP,RefQ}, and the motion
of relativistic projectiles in simpler situations had been studied
earlier \cite{RefR,RefS,RefT}. Similarly, the motion of a relativistic
projectile in a velocity-dependent force field has also been studied
\cite{RefL,RefU}. In the latter references, the constant uniform
external field $\mathbf{F}=-\mathrm{m}g\,\mathrm{\mathbf{\hat{y}}}$
was replaced by $\mathbf{F}=-\mathrm{m}\left(1-\tfrac{v^{2}}{c^{2}}\right)^{-1/2}g\,\mathrm{\mathbf{\hat{y}}}$.
Certainly, the latter choice generates a more realistic model of a
relativistic particle in a uniform gravitational field in an inertial
frame, i.e., within the framework of the special theory of relativity.

References \cite{RefL,RefM,RefN,RefO,RefP,RefQ} did not explicitly
consider the curve that joins the points of maximum height of the
trajectories of the projectile. The principal goal of our work is
to find this curve in the ultra-relativistic limit ($v_{0}\rightarrow c$).
We also want to explicitly verify that the ellipse of eccentricity
$\epsilon=\sqrt{3}/2$ is obtained in the nonrelativistic limit ($v_{0}\ll c$).
To accomplish these two tasks, we first find two exact parametric
equations of the curve that connects the points of maximum height
for all trajectories of a relativistic projectile (here, the parameter
is the launch angle $\theta$). Then, we make the corresponding approximations
on these equations and eliminate $\theta$ in each case to obtain
the locus of the maxima of the trajectories, as shown in Section II.
In Appendix A, using the exact parametric equations of the curve that
passes through the points of maximum height, we calculate the equation
of that curve and its nonrelativistic and ultra-relativistic approximations.
In Appendix B, as a complement to the results in Refs. \cite{RefL,RefN,RefO,RefP,RefQ},
we calculate the exact trajectory of the projectile and its nonrelativistic
and ultra-relativistic approximations. We also obtain the parametric
equations of the trajectory as functions of time in these two approximations.
Additionally, in this section, we find the asymptotic behavior of
the ultra-relativistic trajectory for points that are very distant
from the launching point, i.e., $x\rightarrow\pm\infty$, and the
asymptotic behavior of the two ultra-relativistic parametric equations
for these distant points (in this case, $t\rightarrow\infty$). All
limiting cases that appear throughout the article are carefully studied.
Finally, Section III presents the discussion, which also includes
various physical issues directly connected with some of our results.
For example, we add a numerical estimate of the maximum height that
a relativistic projectile reaches if it is fired vertically from a
place where the gravitational field has a real possibility of being
strong. We show the differences between our results and those obtained
by taking into account the relativistic mass of the projectile and
those provided by general relativity. Because our results apply perfectly
to the problem of a particle of charge $q$ that is fired with an
angle $\theta$ with respect to a horizontal plane in
a transverse uniform electric field, we compare a representative distance
appearing in the relativistic projectile problem in the field $\mathbf{F}=\mathrm{m}\mathbf{g}=\mathrm{\mathbf{const}}$
with those of the electrodynamic problem.

\section{The problem and its solution}

\noindent For a relativistic projectile with a rest mass $\mathrm{m}$
moving in the upper half-plane of the $xy$-plane in a uniform force
field $\mathbf{F}=-F\,\mathrm{\mathbf{\hat{y}}}$, the components
of the equation of motion $\mathbf{F}=\mathrm{d}\mathbf{p}/\mathrm{d}t$
are 
\begin{equation}
F_{x}=\frac{\mathrm{d}}{\mathrm{d}t}\, p_{x}\;\Rightarrow\;0=\frac{\mathrm{d}}{\mathrm{d}t}(\mathrm{m}\gamma v_{x})
\end{equation}
and 
\begin{equation}
F_{y}=\frac{\mathrm{d}}{\mathrm{d}t}\, p_{y}\;\Rightarrow\;-F=\frac{\mathrm{d}}{\mathrm{d}t}(\mathrm{m}\gamma v_{y}),
\end{equation}
where $\gamma\equiv(1-\beta^{2})^{-1/2}$, and $\beta\equiv v/c$,
as usual. The respective potential energy is $\mathrm{m}V(y)=Fy$;
therefore, if $F=\mathrm{m}g$, the gravitational potential is given
by $V(y)=gy\equiv\phi$. We integrate Eqs. (1) and (2) and impose
the following initial conditions:
\begin{equation}
v_{x}(0)=v_{0}\cos(\theta)\quad\Rightarrow\quad p_{x}(0)=\left(\mathrm{m}\gamma v_{x}\right)(0)=\mathrm{m}\gamma_{0}v_{0}\cos(\theta)=p_{0}\cos(\theta),
\end{equation}
and 
\begin{equation}
v_{y}(0)=v_{0}\sin(\theta)\quad\Rightarrow\quad p_{y}(0)=\left(\mathrm{m}\gamma v_{y}\right)(0)=\mathrm{m}\gamma_{0}v_{0}\sin(\theta)=p_{0}\sin(\theta),
\end{equation}
where $\theta\in[0\text{\textdegree},180\text{\textdegree}]$, $\gamma_{0}=(1-\beta_{0}^{2})^{-1/2}$
with $\beta_{0}=v_{0}/c$ ($p_{0}=\gamma_{0}\mathrm{m}v_{0}$ is the
magnitude of the initial momentum of the projectil); then, we obtain
\begin{equation}
\mathrm{m}\gamma v_{x}=p_{0}\cos(\theta)
\end{equation}
and
\begin{equation}
\mathrm{m}\gamma v_{y}=p_{0}\sin(\theta)-Ft.
\end{equation}
Substituting $v_{x}$ and $v_{y}$ from Eqs. (5) and (6) into the
formula 
\begin{equation}
\beta^{2}=\frac{v_{x}^{2}+v_{y}^{2}}{c^{2}}=1-\frac{1}{\gamma^{2}},
\end{equation}
yields the following result:
\begin{equation}
\gamma=\frac{\sqrt{\mathrm{m}^{2}c^{2}+p_{0}^{2}-2Ft\, p_{0}\sin(\theta)+F^{2}\, t^{2}}}{\mathrm{m}c}.
\end{equation}
The components of the velocity vector $\mathbf{v}=\mathrm{d}\mathbf{r}/\mathrm{d}t$
as functions of time are obtained by substituting the latter result
into Eqs. (5) and (6), i.e., 
\begin{equation}
v_{x}(t)=\frac{\mathrm{d}}{\mathrm{d}t}\, x(t)=\frac{cp_{0}\cos(\theta)}{\sqrt{\mathrm{m}^{2}c^{2}+p_{0}^{2}-2Ft\, p_{0}\sin(\theta)+F^{2}\, t^{2}}}
\end{equation}
and
\begin{equation}
v_{y}(t)=\frac{\mathrm{d}}{\mathrm{d}t}\, y(t)=\frac{cp_{0}\sin(\theta)-cFt}{\sqrt{\mathrm{m}^{2}c^{2}+p_{0}^{2}-2Ft\, p_{0}\sin(\theta)+F^{2}\, t^{2}}}.
\end{equation}
If the particle reaches the maximum height of its trajectory at time
$\tau$, we can write $v_{y}(\tau)=0$, from which the following result
is obtained: 
\begin{equation}
\tau=\frac{p_{0}\sin(\theta)}{F}.
\end{equation}
Although the force on the particle lies on the $y$-axis, $v_{x}(t)$
is not constant; however, the $x$-component of the proper velocity
$u_{x}(t)=\gamma v_{x}(t)$ is constant as expected (see Eqs. (8)
and (9)).

Similarly, by integrating Eqs. (9) and (10) and imposing the initial
conditions $x(0)=y(0)=0$, we obtain the components of the position
vector $\mathbf{r}$ as functions of time, i.e.,
\begin{equation}
x(t)=\frac{cp_{0}\cos(\theta)}{F}\ln\left[\frac{\sqrt{E_{0}^{2}-2c^{2}Ft\, p_{0}\sin(\theta)+c^{2}F^{2}t^{2}}-cp_{0}\sin(\theta)+cFt}{E_{0}-cp_{0}\sin(\theta)}\right]
\end{equation}
and 
\begin{equation}
y(t)=\frac{1}{F}\left[E_{0}-\sqrt{E_{0}^{2}-2c^{2}Ft\, p_{0}\sin(\theta)+c^{2}F^{2}t^{2}}\,\right],
\end{equation}
where $E_{0}=\sqrt{c^{2}p_{0}^{2}+\mathrm{m}^{2}c^{4}}$ is the initial
relativistic energy of the projectile. Naturally, when $t=2\tau$,
we obtain $y=0$. Likewise, $v_{x}(2\tau)=v_{0}\cos(\theta)$ and
$v_{y}(2\tau)=-v_{0}\sin(\theta)$. 

The components of the position vector at the points of maximum height
of the trajectories are obtained by evaluating $x(t)$ and $y(t)$
at $t=\tau$, from which the following results are obtained: 
\begin{equation}
x(\tau)=\frac{cp_{0}\cos(\theta)}{2F}\ln\left[\frac{E_{0}+cp_{0}\sin(\theta)}{E_{0}-cp_{0}\sin(\theta)}\right]
\end{equation}
and 
\begin{equation}
y(\tau)=\frac{E_{0}}{F}\left[1-\sqrt{1-\frac{c^{2}p_{0}^{2}}{E_{0}^{2}}\sin^{2}(\theta)}\,\right].
\end{equation}
It is convenient to write the latter two expressions in terms of $\beta_{0}$
(instead of $p_{0}$ and $E_{0}$). Because $E_{0}=\mathrm{m}c^{2}(1-\beta_{0}^{2})^{-1/2}$
and $p_{0}=\mathrm{m}v_{0}(1-\beta_{0}^{2})^{-1/2}$, which implies
that $\beta_{0}=cp_{0}/E_{0}$, one can write the following relations:
\begin{equation}
x(\tau)=\frac{\mathrm{m}c^{2}}{2F}\frac{\beta_{0}}{\sqrt{1-\beta_{0}^{2}}}\cos(\theta)\ln\left[\frac{1+\beta_{0}\sin(\theta)}{1-\beta_{0}\sin(\theta)}\right]
\end{equation}
and 
\begin{equation}
y(\tau)=\frac{\mathrm{m}c^{2}}{F}\frac{1}{\sqrt{1-\beta_{0}^{2}}}\left[1-\sqrt{1-\beta_{0}^{2}\sin^{2}(\theta)}\,\right].
\end{equation}
Naturally, the range of the projectile or the maximum horizontal distance
that the projectile reaches as a function of $\theta$, is given by
$\mathrm{R}=2\, x(\tau)=x(2\tau)$. If $\theta\in[0\text{\textdegree},90\text{\textdegree})$,
then $\mathrm{R}>0$; if $\theta\in(90\text{\textdegree},180\text{\textdegree}]$,
then $\mathrm{R}<0$ (if $\theta=90\text{\textdegree}$, then $\mathrm{R}=0$).
In fact, the magnitude of $\mathrm{R}$ is the range of the projectile.
In the next two subsections, we calculate the nonrelativistic and
ultra-relativistic approximations of Eqs. (16) and (17); from these
results, we obtain in each case the curve that joins the points of
maximum height. In Appendix A, we use Eqs. (16) and (17) to obtain
$x(\tau)$ versus $y(\tau)$ and impose the nonrelativistic and ultra-relativistic
limits on this function. 

\subsection{Nonrelativistic approximation}

\noindent In this case, the initial speed of the projectile is low
compared to the speed of light, i.e., $v_{0}\ll c$, or $\beta_{0}\simeq0$;
consequently, we can write the following approximate formulas: 
\begin{equation}
\frac{1}{\sqrt{1-\beta_{0}^{2}}}\simeq1+\frac{1}{2}\beta_{0}^{2},
\end{equation}
\[
\ln\left[\frac{1+\beta_{0}\sin(\theta)}{1-\beta_{0}\sin(\theta)}\right]=\ln\left[1+\beta_{0}\sin(\theta)\right]-\ln\left[1-\beta_{0}\sin(\theta)\right]
\]
\begin{equation}
\simeq\left[\beta_{0}\sin(\theta)-\frac{1}{2}\beta_{0}^{2}\sin^{2}(\theta)\right]-\left[-\beta_{0}\sin(\theta)-\frac{1}{2}\beta_{0}^{2}\sin^{2}(\theta)\right]=2\beta_{0}\sin(\theta)
\end{equation}
and
\begin{equation}
\sqrt{1-\beta_{0}^{2}\sin^{2}(\theta)}\simeq1-\frac{1}{2}\beta_{0}^{2}\sin^{2}(\theta).
\end{equation}
By substituting the relations in Eqs. (18) and (19) into Eq. (16)
and setting $F=\mathrm{m}g$, we obtain the following result in the
nonrelativistic regime: 
\begin{equation}
x(\tau)=\frac{c^{2}\beta_{0}^{2}}{g}\cos(\theta)\sin(\theta)=\frac{v_{0}^{2}}{g}\cos(\theta)\sin(\theta)=\frac{v_{0}^{2}}{2g}\sin(2\theta).
\end{equation}
The nonrelativistic range of the projectile is $\mathrm{R}_{\mathrm{NR}}=2\, x(\tau)=v_{0}^{2}\sin(2\theta)/g$
and can be maximized by making $\mathrm{d}\mathrm{R}_{\mathrm{NR}}/\mathrm{d}\theta=0$
(or $\mathrm{d}(x(\tau))/\mathrm{d}\theta=0$). We obtain the results
$\mathrm{R}_{\mathrm{NR}}(\theta=45\text{\textdegree})=v_{0}^{2}/g$
and $\mathrm{R}_{\mathrm{NR}}(\theta=135\text{\textdegree})=-v_{0}^{2}/g$
as expected. Similarly, by substituting the relations in Eqs. (18)
and (20) into Eq. (17) and setting $F=\mathrm{m}g$, we obtain
\begin{equation}
y(\tau)=\frac{c^{2}\beta_{0}^{2}}{2g}\sin^{2}(\theta)=\frac{v_{0}^{2}}{2g}\sin^{2}(\theta)=\frac{v_{0}^{2}}{4g}\left[\,1-\cos(2\theta)\right].
\end{equation}
Clearly, from Eqs. (21) and (22), we obtain the equation of an ellipse
centered at point $(0,b)$ with semi-major and semi-minor axes $a$
and $b$, i.e.,
\begin{equation}
\frac{\left(x(\tau)-0\right)^{2}}{a^{2}}+\frac{\left(y(\tau)-b\right)^{2}}{b^{2}}=1,
\end{equation}
where $a=v_{0}^{2}/2g$ and $b=a/2$. The maximum height of the projectile
is coordinate $y(\tau)$ evaluated at $\theta=90\text{\textdegree}$,
which is precisely the length of the semi-major axis $a\,(=2b)$.
Different nonrelativistic initial speeds produce different ellipses,
but they have the same eccentricity: $\epsilon=(\sqrt{a^{2}-b^{2}})/a=\sqrt{3}/2$. 

\subsection{Ultra-relativistic approximation}

\noindent In this case, the initial speed of the projectile is comparable
to the speed of light, i.e., $v_{0}\simeq c$, or $\beta_{0}\simeq1$;
consequently, we can write the following approximate formulas:
\begin{equation}
\frac{\beta_{0}}{\sqrt{1-\beta_{0}^{2}}}\simeq\frac{1}{\sqrt{1-\beta_{0}^{2}}}=\gamma_{0}\simeq\frac{1}{\sqrt{2}}\frac{1}{\sqrt{1-\beta_{0}}},
\end{equation}
\begin{equation}
\ln\left[\frac{1+\beta_{0}\sin(\theta)}{1-\beta_{0}\sin(\theta)}\right]\simeq\ln\left[\frac{1+\sin(\theta)}{1-\sin(\theta)}\right]
\end{equation}
and 
\begin{equation}
\sqrt{1-\beta_{0}^{2}\sin^{2}(\theta)}\simeq\left|\,\cos(\theta)\right|.
\end{equation}
By substituting the relations in Eqs. (24) and (25) into Eq. (16)
and setting $F=\mathrm{m}g$, we obtain the following result in the
ultra-relativistic regime:
\begin{equation}
x(\tau)=(x(\tau))(\theta)=\frac{1}{2}\frac{1}{\sqrt{2}}\frac{c^{2}}{g}\frac{1}{\sqrt{1-\beta_{0}}}\cos(\theta)\ln\left[\frac{1+\sin(\theta)}{1-\sin(\theta)}\right].
\end{equation}
Similarly, by substituting the relations given in Eqs. (24) and (26)
into Eq. (17) and setting $F=\mathrm{m}g$, we obtain
\begin{equation}
y(\tau)=(y(\tau))(\theta)=\frac{1}{\sqrt{2}}\frac{c^{2}}{g}\frac{1}{\sqrt{1-\beta_{0}}}\left(\,1-\left|\,\cos(\theta)\right|\,\right).
\end{equation}
The maximum height of the projectile when $\beta_{0}\simeq1$ is obtained
by setting $\theta=90\text{\textdegree}$ in Eq. (28), i.e., 
\begin{equation}
(y(\tau))(\theta=90\text{\textdegree})=\frac{1}{\sqrt{2}}\frac{c^{2}}{g}\frac{1}{\sqrt{1-\beta_{0}}}\equiv y_{\mathrm{max}}(\beta_{0})=y_{\mathrm{max}}.
\end{equation}
If the projectile could be launched vertically at the speed of light,
then $y_{\mathrm{max}}$ will go to infinity as expected. In addition,
$(x(\tau))(\theta=90\text{\textdegree})=0$ as expected; in fact,
L'H\^{o}pital's rule implies that $\underset{\theta\rightarrow90\text{\textdegree}}{\lim}\,(x(\tau))(\theta)=0$.
Similarly, $(x(\tau))(\theta=0\text{\textdegree})=(x(\tau))(\theta=180\text{\textdegree})=0$
and $(y(\tau))(\theta=0\text{\textdegree})=(y(\tau))(\theta=180\text{\textdegree})=0$,
as expected. The ultra-relativistic range of the projectile is $\mathrm{R}_{\mathrm{UR}}=2\,(x(\tau))(\theta)$
and can be maximized by making $\mathrm{d}\mathrm{R}_{\mathrm{UR}}/\mathrm{d}\theta=0$
(or $\mathrm{d}(x(\tau))(\theta)/\mathrm{d}\theta=0$). Then, the
following transcendental equation is obtained:
\begin{equation}
1-\sin(\theta)\ln\left[\frac{1+\sin(\theta)}{\left|\,\cos(\theta)\right|}\right]=0,
\end{equation}
whose solutions in the interval of $\theta\in[0\text{\textdegree},180\text{\textdegree}]$
are given by $\theta\simeq56.46\text{\textdegree}$ and $\theta\simeq123.54\text{\textdegree}$.
We obtain the results $\mathrm{R}_{\mathrm{UR}}(\theta=56.46\text{\textdegree})\simeq1.3255\, y_{\mathrm{max}}$
and $\mathrm{R}_{\mathrm{UR}}(\theta=123.54\text{\textdegree})\simeq-1.3255\, y_{\mathrm{max}}$.
The angle $\theta=56.46\text{\textdegree}$ precisely maximizes the
length of the trajectory of a nonrelativistic projectile. The latter
result was obtained in Ref. \cite{RefV}, and discussed in Ref. \cite{RefO}.

The locus of the maxima of the paths was obtained by eliminating the
angle $\theta$ between Eqs. (27) and (28). Because the launch angle
varies between $\theta=0\text{\textdegree}$ and $\theta=180\text{\textdegree}$,
we have that $\sin(\theta)=+\sqrt{1-\cos^{2}(\theta)}$. Likewise,
if $\theta\in[0\text{\textdegree},90\text{\textdegree})$, then $\left|\,\cos(\theta)\right|=\cos(\theta)$.
From Eq. (27), $x(\tau)>0$; if $\theta\in(90\text{\textdegree},180\text{\textdegree}]$,
then $\left|\,\cos(\theta)\right|=-\cos(\theta)$ and $x(\tau)<0$
(if $\theta=90\text{\textdegree}$, $x(\tau)=0$). Finally, in the
ultra-relativistic limit, $x(\tau)$ as a function of $y(\tau)$ is
given by
\begin{equation}
x(\tau)=x(\tau)(y(\tau))=\pm\frac{1}{2}\left(\, y_{\mathrm{max}}-y(\tau)\right)\,\ln\left[\,\frac{y_{\mathrm{max}}+\sqrt{\, y_{\mathrm{max}}^{2}-\left(\, y_{\mathrm{max}}-y(\tau)\right)^{2}}}{y_{\mathrm{max}}-\sqrt{\, y_{\mathrm{max}}^{2}-\left(\, y_{\mathrm{max}}-y(\tau)\right)^{2}}}\,\right],
\end{equation}
where $0\leq y(\tau)\leq y_{\mathrm{max}}$. The $x(\tau)$ coordinate
reaches its maximum and minimum values when $y(\tau)\simeq0.4475\, y_{\mathrm{max}}$,
i.e., $(x(\tau))_{\mathrm{max}}=0.6627\, y_{\mathrm{max}}$ and $(x(\tau))_{\mathrm{min}}=-(x(\tau))_{\mathrm{max}}$.
Thus, the dimensions of a maximum height curve depend on $y_{\mathrm{max}}$,
i.e., on $\beta_{0}$ (with $\beta_{0}\rightarrow1$).

We can write the curve in Eq. (31) in polar form. For this purpose,
instead of the launch point $0\,\mathrm{\mathbf{\hat{x}}}+0\,\mathbf{\hat{y}}$,
it is mathematically convenient to take the point $0\,\mathrm{\mathbf{\hat{x}}}+y_{\mathrm{max}}\,\mathbf{\hat{y}}$
as the origin of the curve. Thus, the coordinates of the curve are:
\begin{equation}
X(\tau)=x(\tau)\quad\mathrm{and}\quad Y(\tau)=y(\tau)-y_{\mathrm{max}},
\end{equation}
where $-y_{\mathrm{max}}\leq Y(\tau)\leq0$. To express the equation
of the curve in polar coordinates $R(\tau)$ and $\alpha$, we write
the following customary relations: 
\begin{equation}
X(\tau)=(X(\tau))(\alpha)=(R(\tau))(\alpha)\,\cos(\alpha)\quad\mathrm{and}\quad Y(\tau)=(Y(\tau))(\alpha)=(R(\tau))(\alpha)\,\sin(\alpha).
\end{equation}
The angle $\alpha$ starts at $0\text{\textdegree}$ from the reference
direction, which is defined as the line $y(\tau)=y_{\mathrm{max}}$;
however, because $Y(\tau)\in[-y_{\mathrm{max}},0]$, it follows that
$180\text{\textdegree}\leq\alpha\leq360\text{\textdegree}$. Then,
the polar equation of the curve that joins the points of maximum height
of the projectile in the ultra-relativistic limit is obtained by first
substituting the coordinates $x(\tau)$ and $y(\tau)$ from Eq. (32)
into Eq. (31) and subsequently substituting the relations in Eq. (33)
to the new Eq. (31). The results are as follows: 
\begin{equation}
R(\tau)=(R(\tau))(\alpha)=\frac{y_{\mathrm{max}}}{\left|\,\sin(\alpha)\right|\cosh\left(\frac{\left|\,\cos(\alpha)\right|}{\sin(\alpha)}\right)}.
\end{equation}
Thus, the parametric equations of the limit curve that joins the points
of maximum height of the ultra-relativistic projectile are:
\begin{equation}
(x(\tau))(\alpha)=(R(\tau))(\alpha)\,\cos(\alpha)\quad\mathrm{and}\quad(y(\tau))(\alpha)=(R(\tau))(\alpha)\,\sin(\alpha)+y_{\mathrm{max}}.
\end{equation}
We recall that the distance $(R(\tau))(\alpha)$ is measured from
the top of the curve that joins the points of maximum height, and
the angle $\alpha$ is measured around that point but restricted to
the interval $\alpha\in[180\text{\textdegree},360\text{\textdegree}]$.
For example, when $\alpha=270\text{\textdegree}$, $x(\tau)=y(\tau)=0$;
when $\alpha=180\text{\textdegree}$ or $360\text{\textdegree}$,
$x(\tau)\rightarrow0$ but $y(\tau)\rightarrow y_{\mathrm{max}}$. 

The curve given parametrically in Eq. (35) corresponds to the limit
$\beta_{0}\rightarrow1$ (i.e., $v_{0}\rightarrow c$). In figure
1, we plot this maximum height limit curve (in units of $\tfrac{c^{2}}{g}$)
whose dimensions correspond to $\beta_{0}=0.9999999$; therefore,
$y_{\mathrm{max}}=\sqrt{5,000,000}=2,236.068$ (in units of $\tfrac{c^{2}}{g}$).
Figure 1 shows that the limit curve $(\mathbf{r}(\tau))(\alpha)=((x(\tau))(\alpha),(y(\tau))(\alpha))$
with $180\text{\textdegree}\leq\alpha\leq360\text{\textdegree}$ is
onion-shaped (line blue). In figure 1, if the sector $0\text{\textdegree}<\alpha<180\text{\textdegree}$
is included (line red), a lemniscate-type curve, which resembles the
infinity symbol ($\infty$) in the vertical position or the number
eight ($8$), is generated and centered at the point $(0,y_{\mathrm{max}})$.
Figure 1 also shows the trajectories of the projectile in the ultra-relativistic
limit with $\beta_{0}=0.9999999$ for the --first quadrant-- launch
angles $\theta=\left\{ 15\text{\textdegree},\,30\text{\textdegree},\,45\text{\textdegree},\,56.46\text{\textdegree},\,65\text{\textdegree},\,75\text{\textdegree}\right\} $
and --second quadrant-- launch angles $\theta=\left\{ 105\text{\textdegree},\,115\text{\textdegree},\,123.54\text{\textdegree},\,135\text{\textdegree},\,150\text{\textdegree},\,165\text{\textdegree}\right\} $
(black lines). The latter trajectories are obtained from Eq. (B10). 

\section{Final discussion}

\noindent The points in the plane at the maximum heights of all trajectories
of an ultra-relativistic projectile ($\beta_{0}\rightarrow1$) lie
on an onion-shaped curve. The curve is one of the loops of a lemniscate-type
curve. Given an extremely high initial speed, the parametric equations
of that curve are provided in Eq. (35), the polar equation is provided
in Eq. (34), and the polar angle is restricted to the interval $\alpha\in[180\text{\textdegree},360\text{\textdegree}]$.
We recall that the pole of this polar curve is the point $0\,\mathrm{\mathbf{\hat{x}}}+y_{\mathrm{max}}\,\mathbf{\hat{y}}$,
but the launching point of the projectile is the point $0\,\mathrm{\mathbf{\hat{x}}}+0\,\mathbf{\hat{y}}$.
We found that only the curve that passes through the highest points
of all trajectories in the ultra-relativistic limit can be written
in polar form in a simple manner. If the polar angle of the parametric
equations in Eq. (35) completely covers the plane, i.e., $\alpha\in[0\text{\textdegree},360\text{\textdegree})$,
then a lemniscate-type curve is generated with the center at the point
where the projectile will reach the highest height. To our knowledge,
these results have not been published elsewhere. In general,
in mathematics, a lemniscate is any of the many curves that have the
shape of an infinity symbol. We also verified that the locus of the
maxima of the relativistic projectile trajectories in the nonrelativistic
limit is an ellipse of eccentricity $\epsilon=\sqrt{3}/2$. Additionally,
we calculated in detail the nonrelativistic and ultra-relativistic
limits of the trajectory equation for a relativistic projectile and
the limits of the parametric equations of the trajectory as functions
of time. From the latter, we recovered the trajectory equations in
each limit. We also showed that, in the ultra-relativistic limit,
the trajectory of the projectile for points that are distant from
the launching point is approximately exponential.

We have addressed the problem of a relativistic projectile in a uniform
force field, $F=\mathrm{const}$, within the scheme of the special
theory of relativity. Certainly, our results apply to the problem
of a charged particle (charge $q$) that is fired with an angle $\theta$
with respect to a horizontal plane in a transverse uniform electric
field \cite{RefQ}. In this case, $F=qE>0$, and the replacement $g\rightarrow\tfrac{qE}{\mathrm{m}}$,
or $\tfrac{c^{2}}{g}\rightarrow\tfrac{\mathrm{m}c^{2}}{qE}$, must
be made in our results. Incidentally, in the case of an electron,
the latter replacement would be $9.165\times10^{15}\:\mathrm{m}\rightarrow81.872\:\mathrm{m}$,
and in the case of a proton, it would be $9.165\times10^{15}\:\mathrm{m}\rightarrow1.503\times10^{5}\:\mathrm{m}$
(we used $c=2.998\times10^{8}\:\mathrm{m}/\mathrm{s}$, $g=9.807\:\mathrm{m}/\mathrm{s}^{2}$
and $F=1\times10^{-15}\;\mathrm{N}$, i.e., the magnitude of the electric
field would be $6.242\times10^{4}\:\mathrm{N}/\mathrm{C}$). The distances
involved in the gravitational problem are considerably larger than
those involved in the electrodynamics problem, certainly, assuming
that $g$ is not very large.

In relativistic classical mechanics, the equation of motion with $F=\mathrm{m}g=\mathrm{const}$
does not describe a proper relativistic particle in a uniform gravitational
field. As we know, in a uniform gravitational field, the force on
a particle depends on its speed, i.e., $F=\mathrm{m}\gamma g$ (see,
for example, Refs. \cite{RefL,RefU}). On the other hand, within the
framework of the general theory of relativity, the force is essentially
given by $F=\mathrm{m}\tilde{\gamma}g$, where $\tilde{\gamma}=\left(1-\beta^{2}+\tfrac{2\phi}{c^{2}}\right)^{-1/2}$.
In the latter case, for example, for a vertically fired projectile,
the equation of motion corresponds to the equation of motion for the
special relativity case with the replacement $\mathrm{m}\gamma\rightarrow\mathrm{m}\tilde{\gamma}$,
namely $\tfrac{\mathrm{d}}{\mathrm{d}t}(\mathrm{m}\tilde{\gamma}v_{y})=-\mathrm{m}\tilde{\gamma}g$
(see Ref. \cite{RefW} and reference 2 therein). In any case, the
maximum height reached by a vertically fired relativistic projectile
should increase as the initial speed of the projectile approaches
the speed of light ($\beta_{0}\rightarrow1$). We can compare how
this distance approaches infinity in the three scenarios presented
above. (a) When $F=\mathrm{m}g=\mathrm{const}$, we have $Y_{\mathrm{max}}=\tfrac{c^{2}}{g}(\gamma_{0}-1)$
(this relation is Eq. (17), with $\theta=90\text{\textdegree}$, or
Eq. (A5)). (b) When $F=\mathrm{m}\gamma g$, we have $Y_{\mathrm{max}}=\tfrac{c^{2}}{g}\ln(\gamma_{0})$
(see Eq. (37) in Ref. \cite{RefL}). (c) When $F=\mathrm{m}\tilde{\gamma}g$,
we have $Y_{\mathrm{max}}=\tfrac{c^{2}}{g}\tfrac{1}{2}\beta_{0}^{2}\gamma_{0}^{2}$
(see Eq. (8) in Ref. \cite{RefW}). In the nonrelativistic approximation
($\beta_{0}\simeq0$), these three functions of $\beta_{0}$ are almost
equal. If a value of $\beta_{0}$ very close to 1 is chosen, the projectile
(c) is the farthest from the launch point, behind it is (a), and further
back is (b). Thus, in this case, the result obtained by assuming that
the gravitational force is $F=\mathrm{m}g=\mathrm{const}$ lies between
the result that takes into account the relativistic mass (i.e., $F=\mathrm{m}\gamma g$)
and the result compatible with general relativity (i.e., $F=\mathrm{m}\tilde{\gamma}g$,
where $\tilde{\gamma}=\left(1-\beta^{2}+\tfrac{2gy}{c^{2}}\right)^{-1/2}$). 

We can numerically estimate the maximum height $Y_{\mathrm{max}}$
that a relativistic projectile reaches if it is fired vertically from
a place where the gravitational field has a real possibility of being
very large. For example, we could suppose that the projectile is fired
from a point that is slightly above the event horizon of a black hole.
In this case, $v_{0}\simeq c$ (i.e., $\beta_{0}\simeq1$), and the
Newtonian gravitational acceleration at the event horizon is given
by $g=g_{\mathrm{horizon}}=\tfrac{c^{4}}{4GM}$ (we use the latter
result only to estimate the acceleration of gravity; incidentally,
the result is not valid in general relativity; in fact, $g_{\mathrm{horizon}}\rightarrow\infty$
at all event horizons). For a black hole with the mass of the Sun,
i.e., $M = {M_ \odot } = 1.989 \times {10^{30}}\,{\rm{kg}}$ ($G=6.672\times10^{-11}\:\mathrm{N}\,\mathrm{m}^{2}/\mathrm{kg}^{2}$),
we obtain the result $g_{\mathrm{horizon}}=1.522\times10^{13}\:\mathrm{m}/\mathrm{s}^{2}$
(i.e., $g_{\mathrm{horizon}}$ is $1.552\times10^{12}$ times the
gravitational acceleration of the Earth); hence, we have that $\tfrac{c^{2}}{g_{\mathrm{horizon}}}=5.905\times10^{3}\:\mathrm{m}$.
For a projectile fired with $\beta_{0}=0.9999999$, we obtain the
following results (we consider the same three cases mentioned above):
(a) $Y_{\mathrm{max}}=\tfrac{c^{2}}{g_{\mathrm{horizon}}}(\gamma_{0}-1)=1.320\times10^{7}\:\mathrm{m}$,
i.e., $Y_{\mathrm{max}}=13,200\:\mathrm{km}$. (b) $Y_{\mathrm{max}}=\tfrac{c^{2}}{g_{\mathrm{horizon}}}\ln(\gamma_{0})=4.554\times10^{4}\:\mathrm{m}$,
i.e., $Y_{\mathrm{max}}=45.540\:\mathrm{km}$. (c) $Y_{\mathrm{max}}=\tfrac{c^{2}}{g_{\mathrm{horizon}}}\tfrac{1}{2}\beta_{0}^{2}\gamma_{0}^{2}=1.475\times10^{10}\:\mathrm{m}$,
i.e., $Y_{\mathrm{max}}=14,750,000\:\mathrm{km}$. 

The relativistic mass of a projectile depends on the velocity of the
projectile; hence, the gravitational force on the projectile in a
uniform gravitational field is not constant, i.e., $F=\mathrm{m}\gamma g$.
However, let us suppose that the gravitational force is constant,
i.e., $F=\mathrm{m}g=\mathrm{const}$. Under this assumption, we can
estimate the ratio of the mass to the radius of a stellar object (e.g.,
a dense star) such that a relativistic projectile fired vertically
from its surface reaches a maximum height that is infinitesimally
small compared with the radius of the stellar object. In the latter
case, the pole of the lemniscate that joins the points of maximum
height of the ultra-relativistic projectile may not be at a great
distance from the launching point of the projectile. Let us suppose
that the radius of the stellar body of mass $M$ is $R$. From its
surface, the projectile is fired vertically with speed $v_{0}$ and
reaches a maximum height of $H_{\mathrm{max}}$. The conservation
of energy of the projectile gives us the following relation:
\begin{equation}
-\frac{GM}{R}+c^{2}(\gamma_{0}-1)=-\frac{GM}{R+H_{\mathrm{max}}}.
\end{equation}
Substituting the Newtonian formula that gives the acceleration of
gravity (in this case, on the surface of the stellar body), that is,
$g=GM/R^{2}$, into Eq. (36), we obtain the following result:
\begin{equation}
H_{\mathrm{max}}=\frac{\tfrac{c^{2}}{g}(\gamma_{0}-1)}{1-\frac{1}{R}\tfrac{c^{2}}{g}(\gamma_{0}-1)}=\frac{Y_{\mathrm{max}}}{1-\frac{Y_{\mathrm{max}}}{R}}
\end{equation}
(In the last step, we also use the formula that gives the maximum
height reached by a relativistic projectile fired vertically when
$F=\mathrm{m}g=\mathrm{const}$). Clearly, when $Y_{\mathrm{max}}\ll R$,
i.e., $\tfrac{c^{2}}{g}(\gamma_{0}-1)\ll R$, or $\tfrac{c^{2}}{G}(\gamma_{0}-1)\ll\tfrac{M}{R}$,
we have that $H_{\mathrm{max}}\simeq Y_{\mathrm{max}}$. We select
a projectile with $\beta_{0}=0.9999999$; therefore, $\gamma_{0}-1\simeq\gamma_{0}=2,236.068$.
Thus, the ratio of the mass to the radius of the stellar object must
verify the following relation: 
\begin{equation}
\ensuremath{1.055\times{10^{9}}\:\frac{{M_{\odot}}}{{R_{\odot}}}\ll\frac{M}{R}}
\end{equation}
(${R_ \odot } = 6.963 \times {10^8}\,{\rm{m}}$ is the radius of the
Sun). Clearly, the latter relation could be satisfied only by a very
particular stellar object, one that is very massive and has a small
radius. 

We hope that the content of this article will be of interest and of
use to advanced undergraduate students and those who are interested
in the study of good geometric properties related to projectile motion,
particularly for relativistic projectiles in a uniform force field.
Additionally, several results concerning nonrelativistic and ultra-relativistic
approximations that are attractive and useful in themselves were also
presented. 

\section{Appendix A}

\noindent In relativistic dynamics, the equation of the curve that
is formed by joining the maximum heights of the projectile trajectories,
e.g., $x(\tau)$ as a function of $y(\tau)$, can be explicitly obtained
by eliminating the launch angle $\theta$ of the projectile between
Eqs. (16) and (17). The latter pair of equations can also be written
as follows:
\[
x(\tau)=\frac{1}{2}\frac{\mathrm{m}c^{2}}{F}\frac{1}{\sqrt{1-\beta_{0}^{2}}}\,\beta_{0}\cos(\theta)\ln\left[\frac{1+\sqrt{\beta_{0}^{2}-\beta_{0}^{2}\cos^{2}(\theta)}}{1-\sqrt{\beta_{0}^{2}-\beta_{0}^{2}\cos^{2}(\theta)}}\right]\tag{A1}
\]
and
\[
y(\tau)=\frac{\mathrm{m}c^{2}}{F}\frac{1}{\sqrt{1-\beta_{0}^{2}}}\left[1-\sqrt{1-(\beta_{0}^{2}-\beta_{0}^{2}\cos^{2}(\theta))}\,\right].\tag{A2}
\]
Then, from Eq. (A2), if we write the expressions for $\beta_{0}^{2}-\beta_{0}^{2}\cos^{2}(\theta)$
and for $\beta_{0}\cos(\theta)$ and substitute them into Eq. (A1),
we obtain the following preliminary result:
\[
x(\tau)=\pm\frac{1}{2}\sqrt{(d-y(\tau))^{2}-d^{2}(1-\beta_{0}^{2})}\,\ln\left[\,\frac{d+\sqrt{\, d^{2}-(d-y(\tau))^{2}}}{d-\sqrt{\, d^{2}-(d-y(\tau))^{2}}}\,\right],\tag{A3}
\]
where 
\[
d=d(\beta_{0})\equiv\frac{\mathrm{m}c^{2}}{F}\frac{1}{\sqrt{1-\beta_{0}^{2}}}.\tag{A4}
\]
The maximum height that the relativistic projectile reaches is obtained
by setting $\theta=90\text{\textdegree}$ in Eq. (A2), i.e., 
\[
(y(\tau))(\theta=90\text{\textdegree})=\frac{\mathrm{m}c^{2}}{F}\frac{1}{\sqrt{1-\beta_{0}^{2}}}\left(1-\sqrt{1-\beta_{0}^{2}}\right)=d\left(1-\sqrt{1-\beta_{0}^{2}}\right)\equiv Y_{\mathrm{max}}(\beta_{0})=Y_{\mathrm{max}}\tag{A5}
\]
(we also use the definition in Eq. (A4)). We obtain the following
relation from Eq. (A5): 
\[
d^{2}(1-\beta_{0}^{2})=(d-Y_{\mathrm{max}})^{2}.\tag{A6}
\]
By substituting the latter result into Eq. (A3), we finally obtain
$x(\tau)$ as a function of $y(\tau)$ for a relativistic projectile,
i.e., 
\[
x(\tau)=x(\tau)(y(\tau))=\pm\frac{1}{2}\sqrt{(d-y(\tau))^{2}-(d-Y_{\mathrm{max}})^{2}}\,\ln\left[\,\frac{d+\sqrt{\, d^{2}-(d-y(\tau))^{2}}}{d-\sqrt{\, d^{2}-(d-y(\tau))^{2}}}\,\right].\tag{A7}
\]
Importantly, both $d$ and $Y_{\mathrm{max}}$ depend on the initial
speed of the projectile, i.e., they depend on $\beta_{0}$. Certainly,
the curve in Eq. (A7) is the curve that joins the highest points of
all trajectories of a projectile fired with a (constant) relativistic
initial speed. 

In the nonrelativistic regime, the approximate formulas in Eqs. (18)
and (20) (the latter in the form of $\sqrt{1-\beta_{0}^{2}}\simeq1-\tfrac{1}{2}\beta_{0}^{2}$)
are valid; thus, we can write
\[
d(\beta_{0}\simeq0)=\frac{\mathrm{m}c^{2}}{F}\left(1+\frac{1}{2}\beta_{0}^{2}\right)\simeq\frac{\mathrm{m}c^{2}}{F}=\frac{c^{2}}{g}\tag{A8}
\]
and 
\[
Y_{\mathrm{max}}(\beta_{0}\simeq0)=d(\beta_{0}\simeq0)\left[1-\left(1-\frac{1}{2}\beta_{0}^{2}\right)\right]=\frac{c^{2}}{g}\frac{1}{2}\beta_{0}^{2}=\frac{v_{0}^{2}}{2g}\tag{A9}
\]
(We also use $\beta_{0}=v_{0}/c$ and $F=\mathrm{m}g$). Substituting
the latter pair of relations into Eq. (A7) and developing the differences
of squares in the numerator and denominator within the natural logarithm
function, we obtain 
\[
x(\tau)=\pm\frac{1}{2}\sqrt{\left(\frac{c^{2}}{g}-y(\tau)\right)^{2}-\left(\frac{c^{2}}{g}-\frac{v_{0}^{2}}{2g}\right)^{2}}\,\ln\left[\,\frac{\frac{c^{2}}{g}+\sqrt{\, y(\tau)\left(\frac{2c^{2}}{g}-y(\tau)\right)}}{\frac{c^{2}}{g}-\sqrt{\, y(\tau)\left(\frac{2c^{2}}{g}-y(\tau)\right)}}\,\right].\tag{A10}
\]
We can rewrite and approximate the term with the natural logarithm
as follows:
\[
\ln\left[\,\frac{\frac{c^{2}}{g}+\sqrt{\, y(\tau)\left(\frac{2c^{2}}{g}-y(\tau)\right)}}{\frac{c^{2}}{g}-\sqrt{\, y(\tau)\left(\frac{2c^{2}}{g}-y(\tau)\right)}}\,\right]=\ln\left[\,\frac{1+\sqrt{\,2\frac{y(\tau)}{\frac{c^{2}}{g}}-\left(\frac{y(\tau)}{\frac{c^{2}}{g}}\right)^{2}}}{1-\sqrt{\,2\frac{y(\tau)}{\frac{c^{2}}{g}}-\left(\frac{y(\tau)}{\frac{c^{2}}{g}}\right)^{2}}}\,\right]\simeq\ln\left[\,\frac{1+\sqrt{\,\frac{2g}{c^{2}}y(\tau)}}{1-\sqrt{\,\frac{2g}{c^{2}}y(\tau)}}\,\right].\tag{A11}
\]
In the nonrelativistic regime, the distance $\frac{c^{2}}{g}$ is
much larger than $y(\tau)$; hence, the quadratic term within the
square roots is very small. Now, using a slight variation of the approximate
formula in Eq. (19), the result in Eq. (A11) can be written as follows:
\[
\ln\left[\,\frac{1+\sqrt{\,\frac{2g}{c^{2}}y(\tau)}}{1-\sqrt{\,\frac{2g}{c^{2}}y(\tau)}}\,\right]=\ln\left[\,1+\sqrt{\,\frac{2g}{c^{2}}y(\tau)}\,\right]-\ln\left[\,1-\sqrt{\,\frac{2g}{c^{2}}y(\tau)}\,\right]
\]
\[
\simeq\left[\sqrt{\,\frac{2g}{c^{2}}y(\tau)}-\frac{1}{2}\frac{2g}{c^{2}}y(\tau)\right]-\left[-\sqrt{\,\frac{2g}{c^{2}}y(\tau)}-\frac{1}{2}\frac{2g}{c^{2}}y(\tau)\right]=\sqrt{\,\frac{8g}{c^{2}}y(\tau)}.\tag{A12}
\]
Similarly, the term with the square root that multiplies the term
with the natural logarithm function in Eq. (A10) can be written as
follows:
\[
\sqrt{\left(\frac{c^{2}}{g}-y(\tau)\right)^{2}-\left(\frac{c^{2}}{g}-\frac{v_{0}^{2}}{2g}\right)^{2}}=\sqrt{\, y^{2}(\tau)-\frac{2c^{2}}{g}y(\tau)+\frac{2c^{2}}{g}\frac{v_{0}^{2}}{2g}-\left(\frac{v_{0}^{2}}{2g}\right)^{2}}.\tag{A13}
\]
Substituting the results in Eq. (A11) and Eq. (A12) into Eq. (A10)
and considering the result in Eq. (A13), we obtain 
\[
x(\tau)=\pm\frac{1}{2}\sqrt{\, y^{2}(\tau)-\frac{2c^{2}}{g}y(\tau)+\frac{2c^{2}}{g}\frac{v_{0}^{2}}{2g}-\left(\frac{v_{0}^{2}}{2g}\right)^{2}}\,\sqrt{\,\frac{8g}{c^{2}}y(\tau)}
\]
\[
\pm\frac{1}{2}\sqrt{\,\frac{8g}{c^{2}}y^{3}(\tau)-16\, y^{2}(\tau)+16\,\frac{v_{0}^{2}}{2g}y(\tau)-\frac{8g}{c^{2}}\left(\frac{v_{0}^{2}}{2g}\right)^{2}y(\tau)}
\]
\[
\simeq\pm\frac{1}{2}\sqrt{-16\, y^{2}(\tau)+16\,\frac{v_{0}^{2}}{2g}y(\tau)}\tag{A14}
\]
(we recall that $0\leq y(\tau)\leq Y_{\mathrm{max}}(\beta_{0}\simeq0)=\tfrac{v_{0}^{2}}{2g}$,
and the distance $\frac{c^{2}}{g}$ is much larger than $y(\tau)$
and $\tfrac{v_{0}^{2}}{2g}$, i.e., $\tfrac{g}{c^{2}}y(\tau)\ll1$
and $\tfrac{g}{c^{2}}\tfrac{v_{0}^{2}}{2g}\ll1$). Finally, squaring
the formula in Eq. (A14) and multiplying it by $\left(\tfrac{v_{0}^{2}}{2g}\right)^{-2}$,
we obtain 
\[
\frac{x^{2}(\tau)}{\left(\tfrac{v_{0}^{2}}{2g}\right)^{2}}=-4\frac{y^{2}(\tau)}{\left(\tfrac{v_{0}^{2}}{2g}\right)^{2}}+4\frac{v_{0}^{2}}{2g}\frac{y(\tau)}{\left(\tfrac{v_{0}^{2}}{2g}\right)^{2}}=-\frac{y^{2}(\tau)}{\left(\tfrac{v_{0}^{2}}{4g}\right)^{2}}+2\frac{v_{0}^{2}}{4g}\frac{y(\tau)}{\left(\tfrac{v_{0}^{2}}{4g}\right)^{2}}
\]
\[
=\frac{-y^{2}(\tau)+2\frac{v_{0}^{2}}{4g}y(\tau)}{\left(\tfrac{v_{0}^{2}}{4g}\right)^{2}}=\frac{\left(\tfrac{v_{0}^{2}}{4g}\right)^{2}-y^{2}(\tau)+2\frac{v_{0}^{2}}{4g}y(\tau)-\left(\tfrac{v_{0}^{2}}{4g}\right)^{2}}{\left(\tfrac{v_{0}^{2}}{4g}\right)^{2}}
\]
\[
=\frac{\left(\tfrac{v_{0}^{2}}{4g}\right)^{2}-\left(y(\tau)-\tfrac{v_{0}^{2}}{4g}\right)^{2}}{\left(\tfrac{v_{0}^{2}}{4g}\right)^{2}}.\tag{A15}
\]
The latter is precisely the result in Eq. (23), i.e., in the nonrelativistic
approximation, the points of maximum height of all trajectories of
a projectile launched with the same initial speed form an ellipse
of eccentricity $\epsilon=\sqrt{3}/2$. 

In the ultra-relativistic regime, the approximate formulas in Eqs.
(24) and (26) (the latter in the form of $\sqrt{1-\beta_{0}^{2}}\simeq0$)
are true; hence, we can write the following results: 
\[
d(\beta_{0}\simeq1)=\frac{\mathrm{m}c^{2}}{F}\frac{1}{\sqrt{2}}\frac{1}{\sqrt{1-\beta_{0}}}=\frac{1}{\sqrt{2}}\frac{c^{2}}{g}\frac{1}{\sqrt{1-\beta_{0}}}=y_{\mathrm{max}}\tag{A16}
\]
and 
\[
Y_{\mathrm{max}}(\beta_{0}\simeq1)=d(\beta_{0}\simeq1)\left(1-0\right)=d(\beta_{0}\simeq1)=y_{\mathrm{max}}\tag{A17}
\]
(we also use $F=\mathrm{m}g$). Substituting the latter two relations
into Eq. (A7), we obtain 
\[
x(\tau)=\pm\frac{1}{2}\sqrt{(y_{\mathrm{max}}-y(\tau))^{2}-(y_{\mathrm{max}}-y_{\mathrm{max}})^{2}}\,\ln\left[\,\frac{y_{\mathrm{max}}+\sqrt{\, y_{\mathrm{max}}^{2}-(y_{\mathrm{max}}-y(\tau))^{2}}}{y_{\mathrm{max}}-\sqrt{\, y_{\mathrm{max}}^{2}-(y_{\mathrm{max}}-y(\tau))^{2}}}\,\right].\tag{A18}
\]
This result is identical to the result in Eq. (31) as expected (we
recall that $0\leq y(\tau)\leq Y_{\mathrm{max}}(\beta_{0}\simeq1)=y_{\mathrm{max}}$).
In fact, the approximation in Eq. (A17) enables the limit curve in
Eq. (A18) to be written in a simple polar form (i.e., the form in
Eq. (34)). Actually, as $\beta_{0}$ tends to $1$, $d=\tfrac{\mathrm{m}c^{2}}{F}\gamma_{0}$
(Eq. (A4)) tends to $y_{\mathrm{max}}$ and $Y_{\mathrm{max}}=d-\tfrac{\mathrm{m}c^{2}}{F}=\tfrac{\mathrm{m}c^{2}}{F}(\gamma_{0}-1)$
(Eqs. (A4) and (A5)) also tends to $y_{\mathrm{max}}$ (we recall
that $\beta_{0}$ must be very close to $1$ for $\gamma_{0}$ to
be appreciably greater than $1$). From the latter reasoning, the
result in Eq. (A18) finally arises. In figure 2, we plot the functions
or curves in Eq. (A7) (the --exact relativistic curve)-- (orange lines)
and Eq. (A18) (--its ultra-relativistic limit)-- (red and blue lines)
and compare them near the point where the projectile would reach its
maximum height (in units of $\tfrac{c^{2}}{g}$). Figure 2 also shows
the trajectories of the projectile in the ultra-relativistic limit
for the launch angles of $89.99\text{\textdegree}$ and $90.01\text{\textdegree}$
(Eq. (B10), i.e., for projectiles fired almost vertically. Similar
to figure 1, we select $\beta_{0}=0.9999999$; therefore, $Y_{\mathrm{max}}=2,235.068$
($y_{\mathrm{max}}=2,236.068$). 

\section{Appendix B}

The parametric equations of the trajectory are provided in Eqs. (12)
and (13). Eq. (13) can also be written as follows: 
\[
\sqrt{E_{0}^{2}-2c^{2}Ft\, p_{0}\sin(\theta)+c^{2}F^{2}t^{2}}=E_{0}-Fy.\tag{B1}
\]
Therefore,
\[
cFt-cp_{0}\sin(\theta)=\pm\sqrt{c^{2}p_{0}^{2}\sin^{2}(\theta)-2E_{0}\, Fy+F^{2}y^{2}},\tag{B2}
\]
where the positive sign corresponds to the case of $t>\tau$ and the
negative sign corresponds to the case of $t<\tau$. The latter relation
provides the two solutions of the quadratic equation with time as
the unknown quantity (see Eq. (13)). Substituting the relations in
Eqs. (B1) and (B2) into Eq. (12) and exponentiating on both sides
of the equality, the following relation is obtained: 
\[
\mp\sqrt{c^{2}p_{0}^{2}\sin^{2}(\theta)-2E_{0}\, Fy+F^{2}y^{2}}=E_{0}-Fy-(E_{0}-cp_{0}\sin(\theta))\exp\left(\frac{Fx}{cp_{0}\cos(\theta)}\right).\tag{B3}
\]
Squaring both sides of this expression, after some algebra, we obtain
the equation of the projectile trajectory: 
\[
y(x)=\frac{E_{0}}{F}-\frac{E_{0}}{F}\cosh\left(\frac{Fx}{cp_{0}\cos(\theta)}\right)+\frac{cp_{0}\sin(\theta)}{F}\sinh\left(\frac{Fx}{cp_{0}\cos(\theta)}\right),\tag{B4}
\]
which is precisely the result in Refs. \cite{RefN,RefO}. Note that,
in the latter function, $x$ and $\cos(\theta)$ always have the same
sign when $0\text{\textdegree}\leq\theta\leq180\text{\textdegree}$,
whereas $\sin(\theta)$ never changes sign. More precisely, because
$\cos(180\text{\textdegree}-\theta)=-\cos(\theta)$ and $\sin(180\text{\textdegree}-\theta)=\sin(\theta)$,
$y(x)\equiv y(x,\theta)=y(-x,180\text{\textdegree}-\theta)$ as expected.
Now, let us write $y(x)$ in Eq. (B4) in terms of $\beta_{0}$: 
\[
y(x)=\frac{\mathrm{m}c^{2}}{F}\frac{1}{\sqrt{1-\beta_{0}^{2}}}-\frac{\mathrm{m}c^{2}}{F}\frac{1}{\sqrt{1-\beta_{0}^{2}}}\cosh\left(\frac{x}{\frac{\mathrm{m}c^{2}}{F}\frac{\beta_{0}}{\sqrt{1-\beta_{0}^{2}}}\cos(\theta)}\right)
\]
\[
+\frac{\mathrm{m}c^{2}}{F}\frac{\beta_{0}}{\sqrt{1-\beta_{0}^{2}}}\sin(\theta)\sinh\left(\frac{x}{\frac{\mathrm{m}c^{2}}{F}\frac{\beta_{0}}{\sqrt{1-\beta_{0}^{2}}}\cos(\theta)}\right).\tag{B5}
\]
We emphasize again that the latter is the exact equation of the trajectory
of the relativistic projectile.

In the nonrelativistic regime, the approximate formula in Eq. (18)
is valid; thus, if we substitute this formula into Eq. (B5) we can
write: 
\[
y(x)=\frac{\mathrm{m}c^{2}}{F}\left[\,1-\cosh\left(\frac{x}{\frac{\mathrm{m}c^{2}}{F}\beta_{0}\cos(\theta)}\right)+\beta_{0}\sin(\theta)\,\sinh\left(\frac{x}{\frac{\mathrm{m}c^{2}}{F}\beta_{0}\cos(\theta)}\right)\right].\tag{B6}
\]
The argument of the hyperbolic sine and cosine is a small quantity
in the nonrelativistic regime, i.e., when $v_{0}$ is much smaller
than the speed of light (or $\beta_{0}\simeq0$) and $\left|x\right|$
is not a very large distance. Moreover, $\tfrac{\mathrm{m}c^{2}}{F}=\tfrac{c^{2}}{g}$
is a large distance. Thus, we can use the following approximation:
\[
\cosh\left(\frac{x}{\frac{\mathrm{m}c^{2}}{F}\beta_{0}\cos(\theta)}\right)\simeq1+\frac{1}{2}\frac{F^{2}x^{2}}{(\mathrm{m}c^{2})^{2}\beta_{0}^{2}\cos^{2}(\theta)}=1+\frac{1}{2}\frac{1}{c^{2}}\frac{g^{2}}{v_{0}^{2}\cos^{2}(\theta)}x^{2}\tag{B7}
\]
and 
\[
\sinh\left(\frac{x}{\frac{\mathrm{m}c^{2}}{F}\beta_{0}\cos(\theta)}\right)\simeq\frac{Fx}{\mathrm{m}c^{2}\beta_{0}\cos(\theta)}=\frac{1}{c}\frac{g}{v_{0}\cos(\theta)}x\tag{B8}
\]
(see the comment that follows Eq. (B4)). Substituting these two approximations
into Eq. (B6), we obtain the entire trajectory of the nonrelativistic
projectile, i.e., a parabola:
\[
y(x)=-\frac{g}{2v_{0}^{2}}\sec^{2}(\theta)\, x^{2}+\tan(\theta)\, x=\tan(\theta)\, x\left(1-\frac{x}{\mathrm{R}_{\mathrm{NR}}}\right),\tag{B9}
\]
where $\mathrm{R}_{\mathrm{NR}}=\tfrac{v_{0}^{2}}{g}\,2\cos(\theta)\sin(\theta)=\tfrac{v_{0}^{2}}{g}\sin(2\theta)$
is the range of the projectile as a function of the launch angle in
the nonrelativistic regime (see the discussion following Eq. (21)).

On the other hand, in the ultra-relativistic regime, approximations
in Eq. (24) are valid. Thus, if we use these formulas in Eq. (B5),
and use the definition in Eq. (29) and $\tfrac{\mathrm{m}c^{2}}{F}=\tfrac{c^{2}}{g}$,
we obtain the entire trajectory of the relativistic projectile in
the ultra-relativistic limit as follows:
\[
y(x)=y_{\mathrm{max}}\left[\,1-\cosh\left(\frac{x}{y_{\mathrm{max}}\cos(\theta)}\right)+\sin(\theta)\,\sinh\left(\frac{x}{y_{\mathrm{max}}\cos(\theta)}\right)\right].\tag{B10}
\]
In general, an ultra-relativistic projectile can reach greater distances
than a nonrelativistic projectile, i.e., it moves further away from
the launching point. Thus, we can use the following two formulas which
are valid for large (positive and negative) values of $x$:
\[
\cosh\left(\frac{x}{y_{\mathrm{max}}\cos(\theta)}\right)\simeq\frac{1}{2}\exp\left(\frac{x}{y_{\mathrm{max}}\cos(\theta)}\right)\tag{B11}
\]
and 
\[
\sinh\left(\frac{x}{y_{\mathrm{max}}\cos(\theta)}\right)\simeq\frac{1}{2}\exp\left(\frac{x}{y_{\mathrm{max}}\cos(\theta)}\right)\tag{B12}
\]
(see the comment that follows Eq. (B4)). Finally, substituting these
two approximate formulas into Eq. (B10), we obtain the asymptotic
behavior of $y$ as a function of $x$ when $x\rightarrow\pm\infty$,
i.e., 
\[
y(x)\rightarrow y_{\mathrm{max}}\left[\,1-\frac{1}{2}\left(1-\sin(\theta)\right)\exp\left(\frac{x}{y_{\mathrm{max}}\cos(\theta)}\right)\right],\tag{B13}
\]
where $y_{\mathrm{max}}$ is given in Eq. (29). Given a launch angle
$\theta$, the real trajectory of the particle in the ultra-relativistic
regime approaches the exponential curve in Eq. (B13). For certain
launch angles, the maximum approach between these two curves can only
occur if the projectile moves below the line $y=0$. However, if the
launch angles of the projectile verify the following approximation
from Eq. (B13), a large approach is observed on the line $y=0$: 
\[
y(x=\mathrm{R}_{\mathrm{UR}})=y_{\mathrm{max}}\left[\,1-\frac{1}{2}\left(1-\sin(\theta)\right)\exp\left(\frac{\mathrm{R}_{\mathrm{UR}}}{y_{\mathrm{max}}\cos(\theta)}\right)\right]\simeq0\tag{B14}
\]
(where $\mathrm{R}_{\mathrm{UR}}=2\,(x(\tau))(\theta)$ with the function
$(x(\tau))(\theta)$ in Eq. (27), and $y_{\mathrm{max}}$ is defined
in Eq. (29)). Then, the following approximation is obtained:
\[
y_{\mathrm{max}}\left[\,1-\frac{1}{2}\left(1+\sin(\theta)\right)\,\right]\simeq0.\tag{B15}
\]
For a given $y_{\mathrm{max}}$, this formula is mainly satisfied
for high projectile firing angles (when the term inside the bracket
becomes truly small). In figure 3, we plot the asymptotic curves in
Eq. (B13) (green lines) with the respective projectile trajectories
in the ultra-relativistic limit in Eq. (B10)) (black lines) for the
angles $\theta=65\text{\textdegree}$, $\theta=75\text{\textdegree}$
and $\theta=80\text{\textdegree}$. The maximum height limit curve
is shown in blue. Similar to figures 1 and 2, we set $\beta_{0}=0.9999999$;
therefore, $y_{\mathrm{max}}=2,236.068$ (in units of $\tfrac{c^{2}}{g}$).

Now, let us write the parametric equations of the trajectory in the
nonrelativistic regime. The parametric equations of the trajectory
are in Eqs. (12) and (13). Let us write the latter equations in terms
of $\beta_{0}$. First, the square root term in both equations can
be written as follows (we also use the relation $cp_{0}=\beta_{0}E_{0}$):
\[
\sqrt{E_{0}^{2}-2\beta_{0}E_{0}\sin(\theta)\, cFt+c^{2}F^{2}t^{2}}=E_{0}\,\sqrt{\,1+\frac{c^{2}F^{2}t^{2}-2\beta_{0}E_{0}\sin(\theta)\, cFt}{E_{0}^{2}}}
\]
\[
=\frac{\mathrm{m}c^{2}}{\sqrt{1-\beta_{0}^{2}}}\,\sqrt{\,1-\beta_{0}\sqrt{1-\beta_{0}^{2}}\,2\sin(\theta)\,\frac{Ft}{\mathrm{m}c}+(1-\beta_{0}^{2})\,\frac{F^{2}t^{2}}{\mathrm{m}^{2}c^{2}}}.\tag{B16}
\]
Similarly, the other terms in Eqs. (12) and (13) can be written as
follows:
\[
\frac{cp_{0}\cos(\theta)}{F}=\frac{\mathrm{m}c^{2}}{F}\frac{\beta_{0}}{\sqrt{1-\beta_{0}^{2}}}\cos(\theta),\tag{B17}
\]
\[
E_{0}-cp_{0}\sin(\theta)=\mathrm{m}c^{2}\,\frac{\left(1-\beta_{0}\sin(\theta)\right)}{\sqrt{1-\beta_{0}^{2}}},\tag{B18}
\]
\[
-cp_{0}\sin(\theta)+cFt=\frac{\mathrm{m}c^{2}}{\sqrt{1-\beta_{0}^{2}}}\,\left(-\beta_{0}\sin(\theta)+\sqrt{1-\beta_{0}^{2}}\,\frac{Ft}{\mathrm{m}c}\right)\tag{B19}
\]
and 
\[
\frac{E_{0}}{F}=\frac{\mathrm{m}c^{2}}{F}\frac{1}{\sqrt{1-\beta_{0}^{2}}}.\tag{B20}
\]
Using the approximate formula in Eq. (18) in Eqs. (B16)--(B20) and
substituting the resulting relations into functions $x(t)$ and $y(t)$
(Eqs. (12) and (13)), we obtain: 
\[
x(t)=\frac{\mathrm{m}c^{2}}{F}\beta_{0}\cos(\theta)\ln\left[\frac{\sqrt{\,1-\beta_{0}\,2\sin(\theta)\,\frac{Ft}{\mathrm{m}c}+\frac{F^{2}t^{2}}{\mathrm{m}^{2}c^{2}}}-\beta_{0}\sin(\theta)+\frac{Ft}{\mathrm{m}c}}{1-\beta_{0}\sin(\theta)}\right]\tag{B21}
\]
and
\[
y(t)=\frac{\mathrm{m}c^{2}}{F}\left[1-\sqrt{\,1-\beta_{0}\,2\sin(\theta)\,\frac{Ft}{\mathrm{m}c}+\frac{F^{2}t^{2}}{\mathrm{m}^{2}c^{2}}}\,\right].\tag{B22}
\]
The second and third terms inside the two square roots are very small
(the times in the nonrelativistic approximation are generally small).
Thus, we can use the following approximation:
\[
\sqrt{\,1-\beta_{0}\,2\sin(\theta)\,\frac{Ft}{\mathrm{m}c}+\frac{F^{2}t^{2}}{\mathrm{m}^{2}c^{2}}}\simeq1-\frac{1}{2}\beta_{0}\,2\sin(\theta)\,\frac{Ft}{\mathrm{m}c}+\frac{1}{2}\frac{F^{2}t^{2}}{\mathrm{m}^{2}c^{2}}.\tag{B23}
\]
Substituting this approximation into Eq. (B21), we obtain the following
preliminary result:
\[
x(t)=\frac{\mathrm{m}c^{2}}{F}\beta_{0}\cos(\theta)\ln\left[\frac{1-\beta_{0}\,\sin(\theta)\,\frac{Ft}{\mathrm{m}c}+\frac{1}{2}\frac{F^{2}t^{2}}{\mathrm{m}^{2}c^{2}}-\beta_{0}\sin(\theta)+\frac{Ft}{\mathrm{m}c}}{1-\beta_{0}\sin(\theta)}\right]
\]
\[
=\frac{\mathrm{m}c^{2}}{F}\beta_{0}\cos(\theta)\ln\left[\frac{1-\beta_{0}\sin(\theta)+(1-\beta_{0}\sin(\theta))\frac{Ft}{\mathrm{m}c}+\frac{1}{2}\frac{F^{2}t^{2}}{\mathrm{m}^{2}c^{2}}}{1-\beta_{0}\sin(\theta)}\right]
\]
\[
=\frac{\mathrm{m}c^{2}}{F}\beta_{0}\cos(\theta)\ln\left[1+\frac{Ft}{\mathrm{m}c}+\frac{1}{2}\frac{1}{(1-\beta_{0}\sin(\theta))}\frac{F^{2}t^{2}}{\mathrm{m}^{2}c^{2}}\right]\simeq\frac{\mathrm{m}c^{2}}{F}\beta_{0}\cos(\theta)\ln\left(1+\frac{Ft}{\mathrm{m}c}\right),\tag{B24}
\]
where we also neglect the quadratic term in time within the natural
logarithm function. We can now use the following approximation, which
is valid for small values of $Ft/\mathrm{m}c$:
\[
\ln\left(1+\frac{Ft}{\mathrm{m}c}\right)\simeq\frac{Ft}{\mathrm{m}c}.\tag{B25}
\]
Substituting the latter formula into Eq. (B24), we obtain $x(t)$
in the nonrelativistic regime:
\[
x(t)=\frac{\mathrm{m}c^{2}}{F}\beta_{0}\cos(\theta)\,\frac{Ft}{\mathrm{m}c}=v_{0}\cos(\theta)\, t\tag{B26}
\]
(we also use $\beta_{0}=v_{0}/c$). Similarly, substituting the approximation
in Eq. (B23) into Eq. (B22), we obtain 
\[
y(t)=\frac{\mathrm{m}c^{2}}{F}\left[1-\left(1-\frac{1}{2}\beta_{0}\,2\sin(\theta)\,\frac{Ft}{\mathrm{m}c}+\frac{1}{2}\frac{F^{2}t^{2}}{\mathrm{m}^{2}c^{2}}\right)\,\right]
\]
\[
=\frac{\mathrm{m}c^{2}}{F}\left(\beta_{0}\,\sin(\theta)\,\frac{Ft}{\mathrm{m}c}-\frac{1}{2}\frac{F^{2}t^{2}}{\mathrm{m}^{2}c^{2}}\right)=v_{0}\sin(\theta)\, t-\frac{1}{2}g\, t^{2}\tag{B27}
\]
(we also use $\beta_{0}=v_{0}/c$ and $F=\mathrm{m}g$). Extracting
variable $t$ from Eq. (B26) and substituting it into Eq. (B27) immediately
yield the function $y=y(x)$ given in Eq. (B9).

Let us now write the parametric equations of the trajectory in the
ultra-relativistic regime. First, we write $x(t)$ and $y(t)$ in
terms of $\beta_{0}$ (see Eqs. (12) and (13)). The square root term
in both equations can be written as follows (we also use the relation
$cp_{0}=\beta_{0}E_{0}$):
\[
\sqrt{E_{0}^{2}-2\beta_{0}E_{0}\sin(\theta)\, cFt+c^{2}F^{2}t^{2}}=\sqrt{\left(\frac{\mathrm{m}c^{2}}{\sqrt{1-\beta_{0}^{2}}}\right)^{2}-\mathrm{m}c^{2}\frac{\beta_{0}}{\sqrt{1-\beta_{0}^{2}}}\,2\sin(\theta)\, cFt+c^{2}F^{2}t^{2}}.\tag{B28}
\]
Likewise, the other terms in $x(t)$ and $y(t)$ were written in terms
of $\beta_{0}$ in Eqs. (B17)--(B20). In this regime, the approximations
in Eq. (24) are valid. It is convenient to write these formulas in
terms of the distance $y_{\mathrm{max}}$ defined in Eq. (29), i.e.,
\[
\frac{\beta_{0}}{\sqrt{1-\beta_{0}^{2}}}\simeq\frac{1}{\sqrt{1-\beta_{0}^{2}}}\simeq\frac{1}{\sqrt{2}}\frac{1}{\sqrt{1-\beta_{0}}}=\frac{g}{c^{2}}\, y_{\mathrm{max}}.\tag{B29}
\]
Substituting these approximations into Eq. (B28), and the relations
in Eqs. (B17)--(B20), we obtain the following results: 
\[
\sqrt{E_{0}^{2}-2\beta_{0}E_{0}\sin(\theta)\, cFt+c^{2}F^{2}t^{2}}=\mathrm{m}g\,\sqrt{\, y_{\mathrm{max}}^{2}-y_{\mathrm{max}}\,2\sin(\theta)\, ct+c^{2}t^{2}},\tag{B30}
\]
\[
\frac{cp_{0}\cos(\theta)}{F}=y_{\mathrm{max}}\cos(\theta),\tag{B31}
\]
\[
E_{0}-cp_{0}\sin(\theta)=\mathrm{m}g\, y_{\mathrm{max}}\left(1-\sin(\theta)\right),\tag{B32}
\]
\[
-cp_{0}\sin(\theta)+cFt=\mathrm{m}g\,\left(-y_{\mathrm{max}}\sin(\theta)+ct\right)\tag{B33}
\]
and 
\[
\frac{E_{0}}{F}=y_{\mathrm{max}}\tag{B34}
\]
(we also use $F=\mathrm{m}g$). Substituting the latter formulas into
parametric functions $x(t)$ and $y(t)$ (Eqs. (12) and (13)), we
obtain:
\[
x(t)=y_{\mathrm{max}}\cos(\theta)\ln\left[\frac{\sqrt{\, y_{\mathrm{max}}^{2}-y_{\mathrm{max}}\,2\sin(\theta)\, ct+c^{2}t^{2}}-y_{\mathrm{max}}\sin(\theta)+ct}{y_{\mathrm{max}}\left(1-\sin(\theta)\right)}\right]\tag{B35}
\]
and 
\[
y(t)=y_{\mathrm{max}}-\sqrt{\, y_{\mathrm{max}}^{2}-y_{\mathrm{max}}\,2\sin(\theta)\, ct+c^{2}t^{2}}.\tag{B36}
\]
If the square root in Eq. (B36) is substituted into Eq. (B35) and
if the quadratic equation for variable $t$ from Eq. (B36) is solved,
its solutions are substituted into Eq. (B35), i.e., in the term that
depends on $t$ and is outside the square root, we obtain the trajectory
equation $y=y(x)$ in Eq. (B10). The expression in Eq. (B13) shows
the trajectory followed by the projectile when $x\rightarrow\pm\infty$.
The parametric equations of this trajectory are precisely Eqs. (B35)
and (B36) when $t\rightarrow\infty$. For example, the speed of the
projectile as a function of time, i.e., $v(t)=\sqrt{v_{x}^{2}(t)+v_{y}^{2}(t)}$,
leads to the result $v(t\rightarrow\infty)\rightarrow c$ (see Eqs.
(9) and (10)). Then, the parametric equations in Eqs. (B35) and (B36)
take the following form when $t\rightarrow\infty$:
\[
x(t)\rightarrow y_{\mathrm{max}}\cos(\theta)\ln\left[\frac{\sqrt{c^{2}t^{2}}+ct}{y_{\mathrm{max}}\left(1-\sin(\theta)\right)}\right]=y_{\mathrm{max}}\cos(\theta)\ln\left[\frac{2ct}{y_{\mathrm{max}}\left(1-\sin(\theta)\right)}\right]\tag{B37}
\]
and 
\[
y(t)\rightarrow y_{\mathrm{max}}-\sqrt{c^{2}t^{2}}=y_{\mathrm{max}}-ct.\tag{B38}
\]
By eliminating the time variable between these two relations, we easily
obtain the formula in Eq. (B13).

\section*{Acknowledgement}

\noindent I would like to express my sincere gratitude to Professor Narongrit Maneejiraprakarn, Director of the Institute for Fundamental Study (IF), Naresuan University, Thailand, for giving me the opportunity to research and teach at the IF and for all the support.

\section*{Conflicts of interest}

\noindent The author declares no conflicts of interest.

\pagebreak{}

\begin{center}

\setlength\fboxsep{0pt}
\setlength\fboxrule{0.8pt}
\fbox{\includegraphics[scale=0.6]{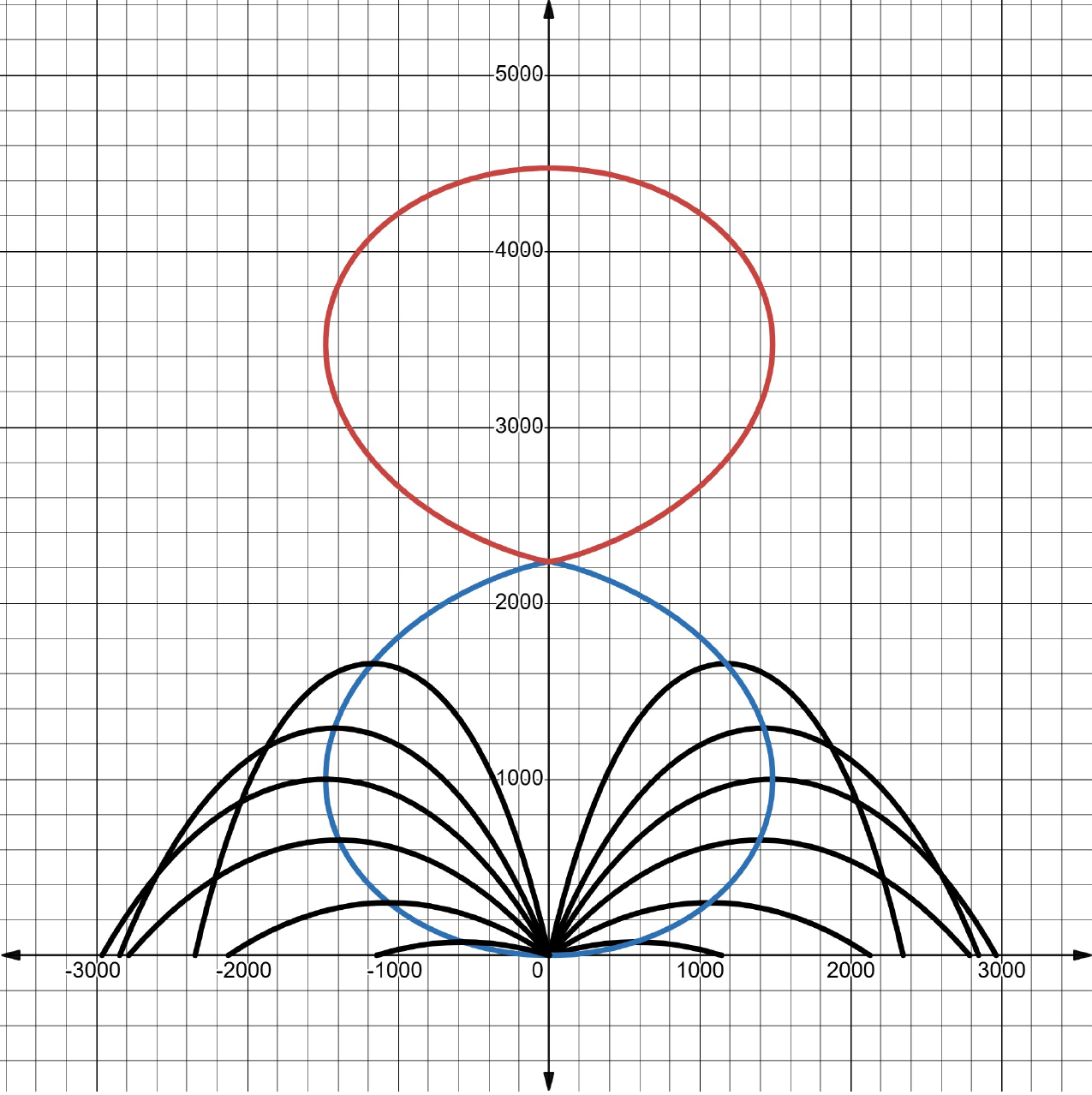}}
\par\end{center}

\noindent \textbf{Figure 1.} Trajectories of a relativistic projectile
in the ultra-relativistic limit for the launch angles $15\text{\textdegree}$,
$30\text{\textdegree}$, $45\text{\textdegree}$, $56.46\text{\textdegree}$,
$65\text{\textdegree}$, $75\text{\textdegree}$, $105\text{\textdegree}$,
$115\text{\textdegree}$, $123.54\text{\textdegree}$, $135\text{\textdegree}$,
$150\text{\textdegree}$ and $165\text{\textdegree}$ (black lines).
An onion-shaped curve joins the points of maximum height of all these
trajectories (blue line). The polar angle of this curve is restricted
to the interval $\alpha\in[180\text{\textdegree},360\text{\textdegree}]$.
However, if the interval $\alpha\in(0\text{\textdegree},180\text{\textdegree})$
is included, a loop similar to the previous one arises (red line).
Clearly, one has a lemniscate-type curve. The parameters are $\beta_{0}=0.9999999$;
therefore, $y_{\mathrm{max}}=\sqrt{5,000,000}=2,236.068$ (in units
of $\tfrac{c^{2}}{g}$).

\pagebreak{}

\begin{center}

\setlength\fboxsep{0pt}
\setlength\fboxrule{0.8pt}
\fbox{\includegraphics[scale=0.6]{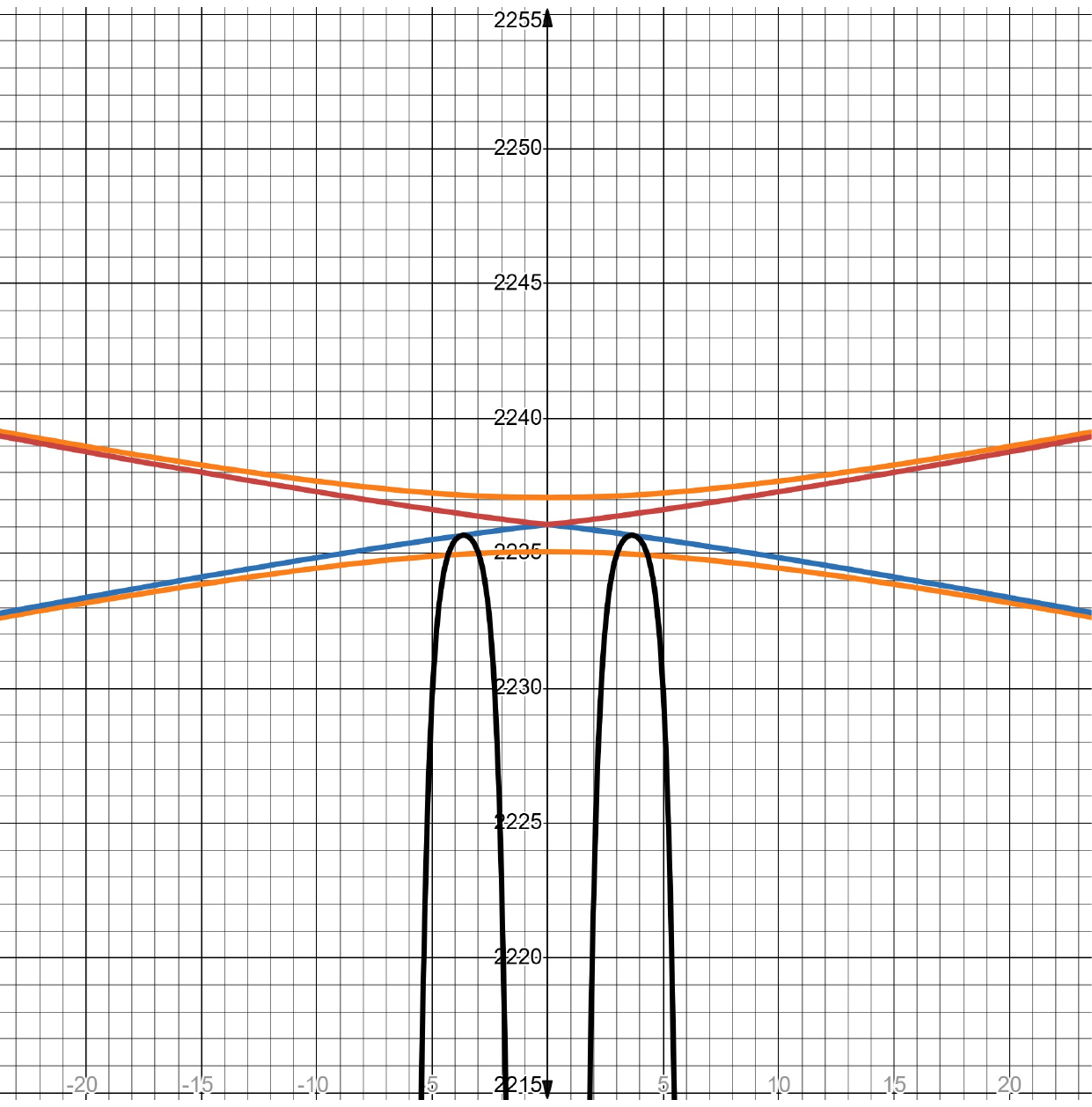}}
\par\end{center}

\noindent \textbf{Figure 2.} Near-pole shot of the lemniscate-type
curve in figure 1. Function $x(\tau)$ versus $y(\tau)$ in Eq. (A7)
is also shown, i.e., the exact locus of the maxima of the paths of
the relativistic projectile (lower orange line). This function also
generates another loop (upper orange line) which is very close to
the lower loop. Trajectories of a relativistic projectile in the ultra-relativistic
limit for the launch angles $89.99\text{\textdegree}$ and $90.01\text{\textdegree}$
(black lines) are also shown.

\pagebreak{}

\begin{center}

\setlength\fboxsep{0pt}
\setlength\fboxrule{0.8pt}
\fbox{\includegraphics[scale=0.6]{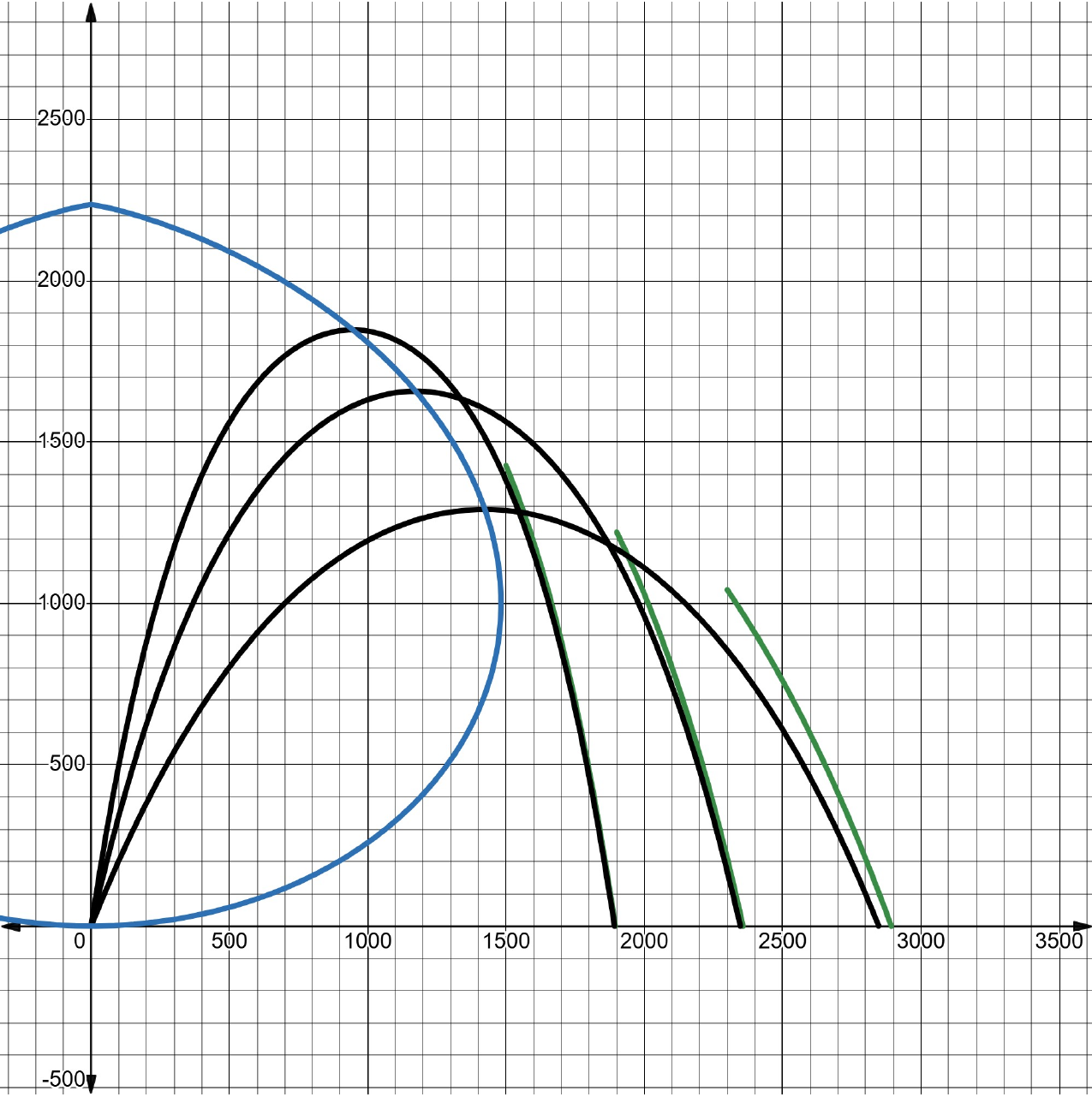}}
\par\end{center}

\noindent \textbf{Figure 3.} Trajectories of a relativistic projectile
in the ultra-relativistic limit for the launch angles $65\text{\textdegree}$,
$75\text{\textdegree}$ and $80\text{\textdegree}$ (black lines)
with the respective asymptotic trajectories of the projectile (green
lines), which are obtained from Eq. (B13). The curve that joins the
points of maximum height of all trajectories is shown in blue. 

\pagebreak{}



\end{document}